# TOPICS ON THE STOCHASTIC TREATMENT OF THE EVOLUTION OF AN OPEN QUANTUM SYSTEM

Ioan Sturzu

"Transilvania" University, Physics Department

Brasov, Eroilor 29, R-2200, Romania

The paper shortly presents the role of Stochastic Processes Theory in the present day Quantum Theory, and the relation to Operational Quantum Physics. The dynamics of an open quantum system is studied on a usual example from Quantum Optics, suggesting the definition of a Neumark-type dilation for the non-thermal states.

**PACS**: 03.65.Ta ; 03.65.Yz ; 05.40.Ca ; 42.50.Lc.

## 1. Introduction

The concept of isolated quantum system was very important in the very beginnings of Quantum Theory. Its usage was related to the classical (strong) form of the individualization principle, which asks for a separate, distinguished existence of every experimentally studied system. However the experiment is a teleological investigation. One has to settle the aim of the investigation, which belongs to some theoretical core. This aim materializes in a procedure of measurements, which also stays as a Bench for the validation of the theoretical core over against the reality. The rules of the comparison of numerical data (empirical sentences) to the theoretical data (predictive sentences) are either independent to the theoretical core (as in Classical Physics), or are given by the theory itself, through the agency of operational definition of the observable (as in Relativity Theory or in Quantum Physics). The later case is, from the ontological point of view a "petitio principii", while - as G. Ludwig have shown [1] - from the epistemological point of view, there exist a way to eluding from this situation, by accepting an intrinsic stochastical character of the genuine quantum phenomenon. In the physical perspective, this is related to a weak interpretation for the individualization principle, compatible with the genuine quantum indistinguishability of the identical micro-particles.

Some conceptual problems still remain debatable: the evolution of quantum systems during measurements, the non-locality of these measurements for correlated, but spatially separated quantum systems, etc. Of course, these problems imply the up-mentioned individualization for quantum systems. Open quantum systems can be studied using the same quantum-theoretical framework, by considering a larger, isolated quantum system (the studied system + the environment). The state of the environment has to be given quantum-mechanically, by stochastical concepts (both classical - for the subjective ignorance of the real microscopically state - and quantum intrinsic types). Consequently, the predictions of the state for the studied system get a stochastically form; the evolution of the system is described by master-type, rather than Liouville equations.

The Heisenberg-picture observables of the system go out from self-adjoinness to the maximally symmetrical case, i.e. from the projectorial measures of the spectral theorem to positive-operator-valued-measures (POVM). Here one has a mathematical correlate of the system+environment image, which is Neumark Theorem; this gives the possibility of extension of the system's Hilbert space $\mathsf{H}$ where a POVM $\{\hat{F}(B)\}_{B\in\boldsymbol{B}}$ is defined, to another one $\tilde{\mathsf{H}}$, endowed with the projectorial measure $\{\hat{E}(B)\}_{B\in\boldsymbol{B}}$, for which the following relation holds:

$$\hat{F}(\mathrm{B})=\hat{\Pi}\hat{E}(\mathrm{B}) \quad \forall B\in \boldsymbol{B}$$

where $\hat{\Pi}$ is the canonical projector $\tilde{\mathsf{H}} \to \boldsymbol{H}$ and $\mathsf{B}$ is the set of real Borelians [2].

## 2. Expectation, random variables, stochastic processes

In the system + environment model for an open quantum system, let $\hat{H}_s$ be the Hamiltonian of the studied system in isolation conditions (its action is in $\mathsf{H}_s$), $\hat{H}_b$ is the Hamiltonian of the environment considered as an isolated system (bath) - it actions in $\boldsymbol{H}_b$, and $\hat{H}_i$ is the interaction Hamiltonian - its action is in the tensor product $\mathsf{H} = \mathsf{H}_s \otimes \mathsf{H}_b$.

The evolution of the whole system is unitary, and given by the total Hamiltonian $\hat{H} = \hat{H}_s \otimes \hat{1} + \hat{1} \otimes \hat{H}_b + \hat{H}_i$ while the state is described by a density operator $\hat{r}$ defined on $\mathbf{H}$.

One can show that $\mathbf{A}$, the algebra of bound selfadjoint operators on $\mathbf{H}$ can be factorized as $\mathbf{A} = \mathbf{A}_0 \otimes \mathbf{C}$ (where $\mathbf{A}_0$ and $\mathbf{C}$ are the subalgebras of, respective, the algebras of all bound self-adjoint operators on $\mathbf{H}_s$ and $\mathbf{H}_b$).

In the non-interactive case, one can force the whole system to enter a factorial state:

$$\hat{r} = \hat{r}_s \otimes \hat{r}_b \tag{1}$$

where $\hat{r}_s$ and $\hat{r}_b$ are density operators on $\mathbf{H}_s$ and $\mathbf{H}_b$. Generally, if $\hat{H}_i \neq 0$, one cannot perform the factorization (1).

In the condition of (1) the application: $\hat{r}_s \mapsto \hat{r}_s \otimes \hat{r}_b$ defined from the predual of $\mathbf{A}_0$ to the product of $\mathbf{A}$ has as adjoint, a norm 1 projector, which can be considered as an expectation, given by $x \otimes y (\in \mathbf{A}) \mapsto j_b(y) \cdot x (\in \mathbf{A}_0)$ where $j_b$ is the application induced on $\mathbf{C}$ by the density operator $\hat{r}_b$.

In the interactive conditions, one can consider the partial trace application on the set of all density operators on $\mathbf{H}$, induced by linearity and continuity from: $\hat{r}_1 \otimes \hat{r}_2 \mapsto tr_b(\hat{r}_2) \cdot \hat{r}_1$. Partial trace is a projector from the predual of $\mathbf{A}$ to the predual of $\mathbf{A}_0$, while its adjoint: $i_0 : \mathbf{A}_0 \to \mathbf{A}$ as an injective *-homomorphism, which can be considered as a noncomutative random variable. Tracing successively at different moments of time, one obtains a stochastic process.

The isolated compound systems evolves unitarily, according to parameter group:

$$\hat{r} \mapsto \hat{U}(t) \cdot \hat{r} \cdot \hat{U}(t)^{+}, t \in \mathbb{R} \tag{2}$$

where $U(t) = \exp\left( \frac{i \cdot t}{\hbar} \cdot \hat{H} \right)$ (This evolution is called Schroedinger image.)

The adjoint of the evolution parameter group is a parameter-group of automorphims which action on $\mathbf{A}$ (the Heinserberg image):

$$\hat{x} \mapsto \hat{U}^{+}(t) \cdot \hat{x} \cdot \hat{U}(t) \qquad (3)$$

The evolution of the studied system is given, in Heinserberg image, by the semigroup $\hat{T}(t)$, actioning on $\mathbf{A}_0$, which is the adjoint of the application:

$$\hat{r}_s \mapsto tr_b(\hat{U}(t)\hat{r}_s \otimes \hat{r}_b \hat{U}^{+}(t)) \qquad (4)$$

Of course, this is meaningful only if there exists a moment when the correlations between the two subsystems can be considered zero. Moreover, one can see that the evolution given by the semigroup $\hat{T}(t)$ cannot, generally, preserve the selfadjointness of the operators from $\mathbf{A}_0$, that is projectorial measures break out to POVM's. If $\hat{E}(x)$ is a projector ( $\hat{E}(x)^2 = \hat{E}(x)$ ), the action of the transition semigroup on it is given by:

$$\hat{T}_t \hat{E}(x) = Tr_b\,((\hat{U}^{t}(t)\hat{E}(x) \otimes \hat{1}_b \hat{U}(1) \cdot (\hat{1}_s \otimes \hat{r}_b))$$

depending on the initial state of the bath. In general, this expression cannot observe the idempotence condition, so it is not anymore a projector [3].

## 3. 1-D optical cavity

The evolution of the type (4) yield to master-type linear equations. If the conditions of preserving the trace and the positivity of the state operator $\hat{r}$ (from now on, we shall drop the index s and the "hats" from operators) are compatible to the following general form for the possible master equations:

$$r = -\frac{1}{\hbar}[H, r] + \sum_n D[C_n]r$$

where $D[C_n]\rho$ are linear applications on the operator set, depending on the bounded operators $C_n$:

$$D[C_n] = c_n r c_n^{+} - \frac{1}{2}(c_n^{+} c_n r + r c_n^{+} c_n) .$$

For an optical 1-D optical cavity, with only one oscillation mod ω, if one of the end mirrors is partially transparent one has:

$$
\begin{aligned}
H &= \hbar w a^{+} a \\
c_1 &= \sqrt{g'} a \\
c_2 &= \sqrt{k} a^{+}
\end{aligned} \qquad (5)
$$

In the ordinary cases γ'>κ.

Using P-representation [4]:

$$r = \int_D d^2 a \cdot P(a, a^*) \,|\, a >< a \,|$$

where |α> are the coherent states , one gets a Focker-Plank equation:

$$\frac{\partial P}{\partial t} = \left[ -iw \left( a^* \frac{\partial}{\partial a^*} - a \frac{\partial}{\partial a} \right) + \frac{1}{2}(g' - k)\left( a \frac{\partial}{\partial a} + a^* \frac{\partial}{\partial a^*} \right) + k \frac{\partial^2}{\partial a \partial a^*} \right] P(a, a^*)$$

which is consistent with the stochastical equation of an Ornstein-Uhlenbeck process:

$$da = -\left( \frac{g - k}{2} + iw \right) a \cdot dt + \sqrt{k}\, dh(t) \qquad (6)$$

where *dh(t)* is a complex Wienner stochastical process:

$$<dh(t)>=<dh^*(t)>=< dh(t)\, dh(t')>= < dh^*(t)\, dh^*(t')>=0$$

$$< dh^*(t)\, dh(t)>=dt$$

The solution of the (6) is:

$$a(t) = e^{-(g+iw)t} \left[ a(0) + \sqrt{k} \int_0^t e^{(g+iw)t'} dh(t') \right] \qquad (7)$$

where it was noted $g = \dfrac{g' - k}{2} > 0$

Computing the variance of the solution:

$$
\begin{aligned}
< a(t), a^*(t) > &= k \cdot e^{-2gt} < \left| \int_0^t e^{(g+iw)t'} dh(t') \right|^2 > \\
&= k e^{-2gt} \lim_{N\to\infty} <| \sum_{n=1}^{N} e^{(g+iw)(n-1)\Delta t} \Delta h_n |^2 >= \\
&= k e^{-2gt} \lim_{N\to\infty} \sum_{n=1}^{N} \sum_{m=1}^{N} e^{(g+iw)(n-1)\Delta t + (g-iw)(m-1)\Delta t} < \Delta h_n \Delta h_m^* >= \\
&= k e^{-2gt} \lim_{N\to\infty} \sum_{n=1}^{N} e^{2g(n-1)\Delta t} \Delta t = k\, e^{-2gt} \int_0^t e^{2gt'} dt = \frac{k}{g}(1 - e^{-gt})
\end{aligned} \qquad (8)
$$

one gets a stationary asymptotic behavior, which is reached, for example, in thermal states.

Conform (6) one has the following solutions for mode $w$ field operators (in Heisenberg image):

$$\hat{a}(t) = e^{-(g+iw)t}[\hat{a}(0) + \sqrt{x}\int_0^t e^{(g+iw)t'} d\hat{h}(t')]$$
$$\hat{a}^+(t) = e^{-(g-iw)t}[\hat{a}^+(0) + \sqrt{x}\int_0^t e^{(g-iw)t'} d\hat{h}^+(t')] \qquad (9)$$

where $d\hat{h}(t')$ and $d\hat{h}^+(t')$ are Wiener-type operatorial stochastical processes (defined in the weak sense):

$$< d\hat{h}(t) \cdot d\hat{h}^+(t) >= dt \cdot \hat{1}$$

Starting from here one can write, also, the position and momentum operators (in Heisenberg image:

$$\hat{Q}(t) = \sqrt{\frac{\hbar}{2mw}}(\hat{a}(t) + \hat{a}^+(t)) = e^{-gt}[\hat{Q} \cdot \cos wt + \frac{1}{m \cdot w} \cdot \hat{P} \cdot \sin wt +$$
$$+ \sqrt{k} \cdot \int_0^t e^{gt'}[\cos w(t'-t) d\hat{h}_1(t') + \sin w(t-t') \cdot d\hat{h}_2(t')]$$
$$\hat{P}(t) = i\sqrt{\frac{m\hbar w}{2}} \cdot (\hat{a}^+(t) - \hat{a}(t)) = e^{-gt} \cdot [\hat{P} \cdot \cos wt - mw \cdot \hat{Q} \cdot \sin wt +$$
$$+ \sqrt{k} \cdot \int_0^t e^{gt'}[d\hat{h}_2(t') \cdot \cos w(t-t') - d\hat{h}_1(t') \cdot \sin w(t-t')]$$

where $d\hat{h}_1(t')$ and $d\hat{h}_2(t')$ are the hermitic operatorial stochastical processes defined by:

$$d\hat{h}(t) = \frac{1}{2} \cdot (d\hat{h}_1(t) + i \cdot d\hat{h}_2(t))$$

## 4. Optical 1-D cavity at equilibrium

The next step is to study this dynamics and the possibility of embedding the studied quantum system in a larger one, for which the evolution is unitary. M. Maasen [5] has studied this possibility in the case when there is dissipation, but the noise term is null (the equilibrium, non-stochastic case). In this case (9) becomes: $a(t) = e^{-(g+iw)t} a(0)$ which corresponds in the complex plane to the transformation:

$$C_t : z \mapsto e^{-(g+iw)t} \cdot z \qquad (10)$$

One will use a particular form of a theorem given by Sz. Nagy [6]:

Given the contraction $C_t : \mathbb{C} \to \mathbb{C}\ z \mapsto e^{-(g+iw)t} \cdot z$ there exist uniquely, up to a unitary equivalence a Hilbert space $\boldsymbol{H}$, a parameter group of unitary operators $U_t : \boldsymbol{H} \to \boldsymbol{H}$, the cyclic vector $v$ and the applications $\boldsymbol{J} : \mathbb{C} \to \boldsymbol{H}$, $\Pi : \boldsymbol{H} \to \mathbb{C}$, for which the following diagram is commutative:

$$\begin{array}{ccc} \mathbb{C} & \xrightarrow{C_t} & \mathbb{C} \\ \boldsymbol{J} \downarrow & & \uparrow \Pi \\ \boldsymbol{H} & \xrightarrow{U_t} & \boldsymbol{H} \end{array}$$

$$\Pi(U_t(\boldsymbol{J}(z)) = C_t(z)$$

The structure $(\boldsymbol{H}, \boldsymbol{J}, \Pi, U_t)$ is named the minimal dilation for $C_t$. One of its realizations is given by:

$$\begin{aligned} &\boldsymbol{H} = L^2(\mathbb{R}, 2g\,dx) \\ &[U_t(u)](x) = u(x+t) \\ &v(x) = \begin{cases} e^{-(g+iw)x}, & x \geq 0 \\ 0, & x < 0 \end{cases} \\ &\boldsymbol{J}(z) = z \cdot v \\ &\Pi(u) = \langle v, u \rangle \end{aligned}$$

One will make use of the vacuum representations of canonical commutation relations (CCR) on the Hilbert space $\mathbf{H}$ in the special case $\boldsymbol{K} = \mathbb{C}$.

$$\hat{W}(f) \cdot \hat{W}(g) = e^{-i \operatorname{Im} \langle f, g \rangle} \hat{W}(f+g) \quad \forall f, g \in \boldsymbol{K}$$

$$z \mapsto \hat{W}(z) = e^{z\hat{a} - z^* \hat{a}^+} \tag{11}$$

Let $\boldsymbol{W}_0(\boldsymbol{K})$ the linear covering of the operators: $\{\hat{W}(f)\}_{f \in \boldsymbol{K}}$ and $\boldsymbol{W}(\boldsymbol{K})$ the strong closeness of $\boldsymbol{W}_0(\boldsymbol{K})$. If $C$ is a contraction $\boldsymbol{K}_1 \to \boldsymbol{K}_2$, one defines the application $\boldsymbol{W}_0(C)$: $\boldsymbol{W}_0(\boldsymbol{K}_1) \to \boldsymbol{W}_0(\boldsymbol{K}_2)$ by:

$$\boldsymbol{W}_0(C)(\hat{W}(f)) = e^{\frac{1}{2}(\|f\|^2 - \|Cf\|^2)} \hat{W}(Cf)$$

which can be strongly-continuously extended to an $\boldsymbol{W}(C)$ defined on $\mathbf{W}(\mathbf{K})$.]

Using (11) one can obtain the transition operator (4) as:

$$\begin{aligned} &T_t = \boldsymbol{W}(C_t) \\ &T_t(\hat{W}(z)) = e^{\frac{1}{2}(e^{-2gt} - 1)|z|^2} \cdot \hat{W}(e^{-(g+iw)t} \cdot z) \end{aligned}$$

which gives:

$$\hat{a}(t) = T_t(\hat{a}) = \frac{\partial}{\partial z^*} T_t(\overset{\wedge}{W}(z))|_{z=0} =$$

$$= \frac{\partial}{\partial z^*} \exp(e^{-(g+iw)t} z\hat{a} - e^{-(g-iw)t} z^* \hat{a})|_{z=0} = e^{-(g+iw)} \hat{a}(0)$$

$$\hat{a}^+(t) = T_t(\hat{a}^+) = -\frac{\partial}{\partial z^*} T_t(\overset{\wedge}{W}(z))|_{z=0} = e^{-(g-iw)} \hat{a}^+(0)$$

that is, the evolution equations for equilibrium optical cavity were recovered.

The case of a thermal optical cavity (that is a cavity at thermal equilibrium with the environment) is also studied in literature [7,8,9].

## 5. Conclusions and developments

In this paper the role of Stochastic Processes Theory in the present day Quantum Theory and its relation to Operational Quantum Physics was presented. The dynamics of an open quantum system was studied on a usual example from Quantum Optics.

An alternate language for the method from section 4 is that of generalized scalar operators, as presented in [10] and references therein. If the state equilibrium is not reached, or the equilibrium approximation is not sufficient, one has to consider in (10) the full expression of (7), that is:

$$C_t : z \mapsto e^{-(g+iw)t} \cdot (z + z(t)) \tag{12}$$

where $z(t) = \sqrt{k} \int_0^t e^{(g+iw)t'} dh(t')$ is a Brownian stochastical process. However, Sz. Nagy theorem is not anymore applicable, because (12) is not a pure (afine) contraction. An open question is that if (12) is, however, similar to a contraction, in the language of [10].